Rethinking conditioning in polarimetry: a new framework beyond $\ell^2$-based metrics

Yuxi Cai[†], An Aloysius Wang[†], Chao He[*]

Department of Engineering Science, University of Oxford, Parks Road, Oxford, OX1 3PJ, UK
[†]These authors contributed equally to this work
*Corresponding author: chao.he@eng.ox.ac.uk

**Abstract**

A standard procedure to achieve accurate, precise, and fast polarization measurement is to choose analyzing and generating polarization states that yield an $\ell^2$-condition number optimized instrument matrix. This strategy works well for rotating-waveplate systems, where the accessible polarization states trace a curve on the Poincaré sphere and the corresponding optimization problem is generally well posed. However, it becomes degenerate for liquid-crystal-based systems, which can generate arbitrary polarization states, and whose additional degrees of freedom allow the optimization of metrics beyond the $\ell^2$-condition number. Leveraging this unique advantage of liquid-crystal polarimeters, we introduce additional performance measures derived from alternative norms and error distributions computed via Monte Carlo simulations to inform the design of measurement schemes. We then experimentally demonstrate their effectiveness in suppressing errors, paving the way for more robust and efficient polarization measurements.

**Introduction**

Polarization is a fundamental property of light [1], and its measurement [2] [3] [4] [5] [6] [7] plays a critical role in applications such as biomedical imaging [8] [9] [10] [11] [12] [13] optical communications [14] [15] [16], sensing [17] [18], and material characterization [19]. Accuracy and speed are the two key aspects of polarization measurements, with speed being especially critical in industrial settings where high-throughput sample characterization is required, and in biomedical applications such as polarization microscopy [20] [21] [22], whole-slide imaging [23] and real-time endoscopic polarization measurement [24] [25]. As such, maintaining accuracy while taking only the minimal number of measurements required for Stokes or Mueller matrix reconstruction—4 and 16, respectively—is an important and well-studied problem in polarization imaging and analysis.

The standard approach to suppressing errors is to ensure that the instrument matrix is well conditioned, with optimization of the $\ell^2$-condition number subject to experimental constraints being the usual procedure. For example, a well-known result [26] [27] [28] [29] [30] shows that this condition number is optimized when the analyzing states form a regular tetrahedron (Fig. 1), which can be experimentally realized with a rotating 132° wave plate [29]. Fundamentally, a small condition number is ideal for two key reasons: counteracting measurement noise and mitigating modeling error. To understand this mathematically, consider a Stokes measurement with true instrument matrix $\Sigma_{\text{true}}$ and incident polarization state $s$. The measured intensity is subject to noise, so that

$$I_{\text{measured}} = \Sigma_{\text{true}} s + \delta I$$

where $\delta I$ represents sensor noise. If $\Sigma_{\text{model}} = \Sigma_{\text{true}} + \delta\Sigma$ is the modelled instrument matrix, then the reconstructed polarization state in the case where only four measurements are used is

$$\hat{s} = \Sigma_{\text{model}}^{-1} I_{\text{measured}}.$$

The Taylor expansion $(\Sigma_{\text{true}} + \delta\Sigma)^{-1} = \Sigma_{\text{true}}^{-1} - \Sigma_{\text{true}}^{-1} \delta\Sigma \Sigma_{\text{true}}^{-1} + \mathcal{O}(\|\delta\Sigma\|^2)$ then gives

$$\begin{aligned}\|s - \hat{s}\| &= \|-\Sigma_{\text{true}}^{-1} \delta\Sigma s + \Sigma_{\text{true}}^{-1} \delta I + \mathcal{O}(\|\delta\Sigma\|^2 + \|\delta\Sigma\|\|\delta I\| + \|\delta I\|^2)\| \\ &\leq \|\Sigma_{\text{true}}^{-1} \delta\Sigma s\| + \|\Sigma_{\text{true}}^{-1} \delta I\| + \mathcal{O}(\|\delta\Sigma\|^2 + \|\delta\Sigma\|\|\delta I\| + \|\delta I\|^2)\end{aligned}$$

by repeated application of the triangle inequality. This then yields an estimate, accurate up to first order, given by

$$\frac{\|s - \hat{s}\|}{\|s\|} \leq \kappa(\Sigma_{\text{true}}) \left( \frac{\|\delta\Sigma\|}{\|\Sigma_{\text{true}}\|} + \frac{\|\delta I\|}{\|I_{\text{true}}\|} \right)$$

where $\kappa(\Sigma_{\text{true}}) = \|\Sigma_{\text{true}}\|\|\Sigma_{\text{true}}^{-1}\|$ denotes the condition number of the matrix $\Sigma_{\text{true}}$ and $I_{\text{true}} = \Sigma_{\text{true}} s$. As such, the condition number represents a bound on the worst-case error arising from modelling inaccuracies and measurement noise, and ensuring that it is as small as possible improves the overall robustness of the system.

There are, however, several caveats to using the $\ell^2$-condition number as the sole figure of merit in designing measurement schemes. First, the condition number, being a worst-case bound, governs only the tails of the error distribution and does not control the entry-wise distribution of errors. Consequently, even in the $\ell^2$-optimized setting, there are no guarantees regarding how error is apportioned across individual components, and it may be disproportionately concentrated in a single component.

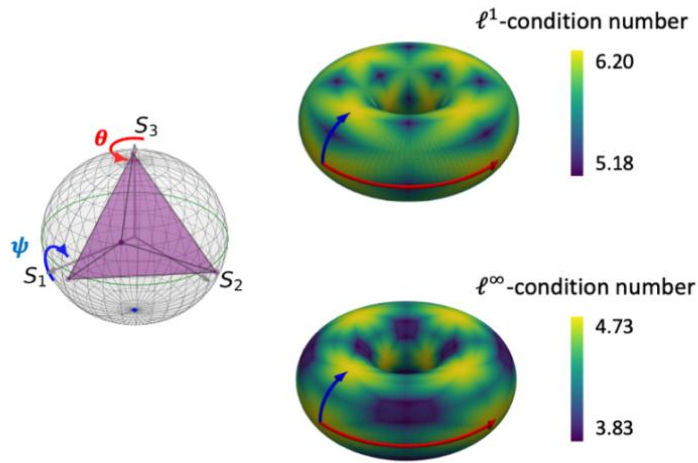

Fig. 1: Concept. When the polarization analysis states of the instrument matrix form a regular tetrahedron within the Poincaré sphere, the resulting measurement strategy achieves the optimal $\ell^2$-condition number. However, condition numbers with respect to different norms may differ. The torus depicts variation in $\ell^1$- and $\ell^\infty$-condition numbers as a fixed tetrahedron is rotated first by angle $\theta$ about the $s_3$-axis followed by angle $\psi$ about the $s_1$-axis.

Second, the optimality condition derived in [29] and [30] implies that there exist infinitely many instrument matrices that are $\ell^2$-condition number optimized, each associated with a distinct error distribution and potentially different $\ell^p$-condition numbers for $p \neq 2$. For example, as demonstrated in Fig. 1, rotating any fixed regular tetrahedron about different axes produces measurement schemes that have different $\ell^1$- and $\ell^\infty$-condition numbers while retaining $\ell^2$-optimality. Thus, a more refined metric is required to identify the most favourable measurement scheme.

Third, the robustness bound derived above assumes model disturbances of an arbitrary linear form, whereas in practice the admissible perturbations $\delta\Sigma$ typically lie in a restricted subset of all matrices. For example, angular misalignments or retardance deviations in the experimental setup constitute inherently non-linear constraints, and their effects can be accounted for through careful modelling. As such, even if the condition number is optimized, it need not be optimal with respect to the subset of physically realizable model errors.

Before addressing these three caveats in detail, it is useful to highlight an additional practical consideration in designing measurement schemes arising from experimental constraints. Notably, in traditional rotating-waveplate systems [2], the accessible generating and analysing polarization states lie on a curve on the Poincaré sphere, so not every state is realizable. To address this limitation, we consider liquid-crystal-based polarimeters (Supplementary Note 1) which, in addition to operating orders of magnitude faster [31], is capable of producing arbitrary polarization states on the Poincaré sphere.

Two main strategies are now introduced to address the caveats outlined above, with the discussion focusing primarily on Stokes measurements for clarity. It should be noted, however, that these strategies are potentially applicable to Mueller matrix measurements as well. The first strategy is to optimize condition numbers under different norms while maintaining optimality of the $\ell^2$-condition number. In addition to being computationally inexpensive, the motivation for this approach is that suppressing error with respect to multiple norms is likely to suppress the overall error. Of particular relevance is the $\ell^\infty$-norm, whose suppression provides robustness guarantees across all entries, thereby addressing the first caveat introduced above.

The second strategy is to use Monte Carlo simulations (Supplementary Note 2) to directly compute the error distribution under sensor noise and modelling uncertainties. This approach not only accommodates non-linear constraints, such as the effects of angular errors on the instrument matrix, thereby addressing the third caveat, but also provides entry-wise error information that can further guide design choices.

Fig. 2a shows the $\ell^1$-, $\ell^2$- and $\ell^\infty$-condition numbers of the instrument matrix as a fixed regular tetrahedron is rotated about the $s_3$-axis, together with the mean absolute errors of each component of the reconstructed Stokes vector, and overall error across all components. The various errors were estimated using Monte Carlo simulations, as detailed in Supplementary Note 2. Notice from the figure that there is a clear link between the two proposed strategies, where the point of overall smallest error also corresponds to a situation where the $\ell^1$- and $\ell^\infty$-condition numbers are optimized. Additionally, note that although the $\ell^2$-condition number is optimized in each case, the error distribution across the various entries can vary substantially, with the error in some components exceeding that of others by more than a factor of two in certain configurations.

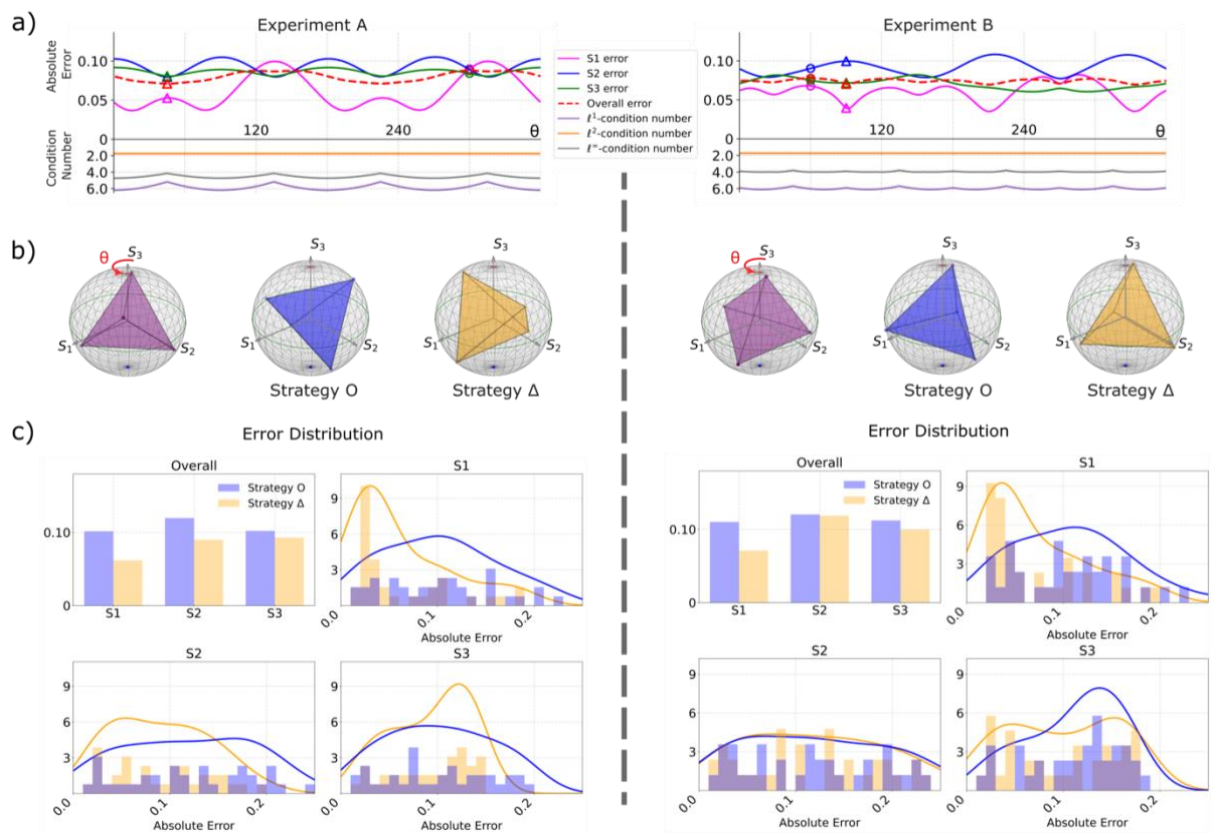

Fig. 2: Error behavior and distribution across the three Stokes parameters. a) Error distributions and condition numbers for measurement configurations obtained by rotating a fixed regular tetrahedron about the $s_3$-axis. b) Visualisations of the corresponding measurement strategies on the Poincaré sphere are shown, with the strategies yielding the best (strategy Δ) and worst (strategy O) overall errors highlighted. c) Experimental results for the selected configurations, showing the error distributions of the different Stokes parameters along with their corresponding histograms. The experimentally computed mean absolute error for each component is also shown.

Lastly, Fig. 2c shows the experimentally measured error distribution of Stokes measurements, obtained using a calibrated polarization state generator (Supplementary Note 1) to produce 200 polarization states distributed uniformly over the Poincaré sphere according to a Fibonacci-spiral scheme. Note that the polarization state analyzer is intentionally left uncalibrated to demonstrate how different measurement schemes can be used to suppress errors. In each experiment, the two measurement strategies employed correspond to those yielding the best and worst overall errors as computed in Fig. 2a. The resulting measured error distributions are qualitatively consistent with the Monte Carlo simulations, further underscoring the usefulness of this approach for understanding and designing measurement schemes. Moreover, the figure clearly shows that the error distribution in the best case outperforming that of the worst case, highlighting that $\ell^2$-optimality is insufficient as a criterion for identifying the best possible measurement scheme.

**Discussion**

Apart from addressing the challenges associated with designing optimized measurement schemes, the experiments and simulations presented in this paper have several important implications for polarization measurement. First, they highlight a key advantage of using liquid-crystal polarimeters that extends beyond speed, namely that the flexibility to produce arbitrary generation and analysis channels enables the design of more robust systems that are less sensitive to modelling errors and sensor noise.

Second, our work demonstrates that different measurement schemes can exhibit large variations in their accuracy when measuring specific components, even when they are $\ell^2$-condition number optimized. This is particularly relevant for systems in which one Stokes parameter is more important than the others, in which case our Monte-Carlo-based strategy can be adapted to generate measurement schemes optimized for that specific component. For example, certain biological and chemical systems exhibit circular dichroism and other chiral responses, and thus behave differently under left- and right-circular polarization, making $s_3$ the most relevant component [32]. On the other hand, in stress and birefringence imaging [33] [34] as well as in many communication systems [35], it is typically linearly polarized light that is most important. Lastly, for applications in structured light, such as those involving complex vector beams [36] and Stokes skyrmions [37] [38], where all polarization states can appear, it is ideal for the error to be uniformly suppressed across all components. Therefore, numerically computed error distributions can serve as an invaluable tool for optimizing polarization measurements in a manner tailored to each specific application.

The point above also extends to Mueller matrix imaging, where, depending on the application, different elements of the measured Mueller matrix may be the most relevant [39] [40] [41] [42]. For instance, in certain cancer-related pathological analysis scenarios, the polarimetric measurement precision of circular polarization (mainly related to the lower submatrix of the Mueller matrix) is more important than its linear counterpart [8] [43], while linear polarization (mainly related to the upper 3×3 submatrix of the Mueller matrix) is given more consideration in certain fluorescence information–based applications [44].

Third, we note that the strategies we have introduced are not limited to cases where exactly 4 measurements are taken in Stokes imaging or 16 in Mueller matrix imaging. In cases where not every component is of interest, fewer measurements may be taken, whereas more measurements can be acquired if a more robust reconstruction is required. In either situation, the Moore-Penrose inverse can be used to recover the desired quantities, and a similar analysis of condition numbers and error distributions can be used to inform the design choice. We leave a more detailed exploration of these extensions to future work.

To conclude, we have demonstrated that relying solely on the $\ell^2$-condition number as the figure of merit is insufficient for optimizing polarization measurement schemes. To address this, we proposed two alternative strategies: one that is computationally simple but does not reveal the distribution of errors among components, and another that incorporates non-linear considerations and provides error distributions at the cost of greater computational effort. We go on to experimentally demonstrate that measurement schemes derived in this way can suppress errors, thereby validating our theory. Taken together, these contributions offer practical pathways for addressing key engineering challenges in the design and optimization of fast yet accurate polarization measurement systems.

# Bibliography

[1]   D. H. Goldstein, "Polarized light," *CRC press,* 2017.

[2]   R. Azzam, "Stokes-vector and Mueller-matrix polarimetry.," *Journal of the Optical Society of America,* 2016.

[3]   R. Azzam, "Photopolarimetric measurement of the Mueller matrix by Fourier analysis of a single detected signal," *Optics Letters,* 1978.

[4]   A. Zaidi et al., "Metasurface-enabled single-shot and complete Mueller matrix imaging," *Nature Photonics,* 2024.

[5] C. He et al., "Full Poincaré polarimetry enabled through physical inference," *Optica,* 2022.

[6] C. He et al., "A Stokes polarimeter based on four quadrant detector," *Journal of Infrared and Millimeter Waves,* 2016.

[7] F. Snik et al., "An overview of polarimetric sensing techniques and technology with applications to different research fields," *Polarization: measurement, analysis, and remote sensing XI,* 2014.

[8] C. He et al., "Polarisation optics for biomedical and clinical applications: a review," *Light Sci Appl,* 2021.

[9] H. He et al., "Mueller matrix polarimetry—an emerging new tool for characterizing the microstructural feature of complex biological specimen," *Journal of Lightwave Technology,* 2018.

[10] C. He et al., "Characterizing microstructures of cancerous tissues using multispectral transformed Mueller matrix polarization parameters," *Biomedical optics express,* 2015.

[11] E. Du et al., "Mueller matrix polarimetry for differentiating characteristic features of cancerous tissues.," *Journal of biomedical optics,* 2015.

[12] J. Fan et al., "Stain transformation using Mueller matrix guided generative adversarial networks," *Optics Letters,* 2024.

[13] L. Deng et al., "Influence of hematoxylin and eosin staining on linear birefringence measurement of fibrous tissue structures in polarization microscopy," *Journal of Biomedical Optics,* 2023.

[14] A. A. Wang et al., "Topological protection of optical skyrmions through complex media," *Light Sci Appl,* 2024.

[15] H. Yan and G. Li., "Coherent optical communication using polarization multiple-input-multiple-output.," *Optics Express,* 2005.

[16] A. A. Wang et al., "Optical Skyrmions in Waveguides," *arXiv preprint arXiv:2505.06735,* 2025.

[17] Y. Ma et al., "Using optical skyrmions to assess vectorial adaptive optics capabilities in the presence of complex aberrations," *Sci. Adv.,* 2025.

[18] S. Cloude, "Polarisation: applications in remote sensing," *Oxford university press,* 2010.

[19] S. Liu, X. Chen and C. Zhang, "Development of a broadband Mueller matrix ellipsometer as a powerful tool for nanostructure metrology," *Thin Solid Films,* 2015.

[20] Y. Ma et al., "Vectorial adaptive optics for advanced imaging systems," *Journal of Optics,* 2024.

[21] C. He, J. Antonello and M. J. Booth, "Vectorial adaptive optics," *eLight,* 2023.

[22] S. Shen et al., "Polarization Aberrations in High-Numerical-Aperture Lens Systems and Their Effects on Vectorial-Information Sensing," *Remote Sensing,* 2022.

[23] Y. Du et al., "Beyond H&E: Unlocking Pathological Insights with Polarization via Self-supervised Learning," *arXiv preprint arXiv:2503.05933,* 2025.

[24] J. Qi et al, "Surgical polarimetric endoscopy for the detection of laryngeal cancer," *Nat. Biomed. Eng,* 2023.

[25] Z. Zhang et al., "Analysis and optimization of aberration induced by oblique incidence for in-vivo tissue polarimetry," *Optics Letters,* 2023.

[26] A. Peinado et al., "Optimization and performance criteria of a Stokes polarimeter based on two variable retarders," *Opt. Express,* 2010.

[27] J. S. Tyo, "Design of optimal polarimeters: maximization of signal-to-noise ratio and minimization of systematic error," *Applied optics,* 2002.

[28] J. S. Tyo, "Noise equalization in Stokes parameter images obtained by use of variable-retardance polarimeters," *Optics Letters,* 2000.

[29] D. Sabatke et al., "Optimization of retardance for a complete Stokes polarimeter," *Opt. Lett,* 2000.

[30] M. R. Foreman, A. Favaro and A. Aiello, "Optimal frames for polarisation state reconstruction.," *Physical Review Letters,* 2015.

[31] J. M. López-Téllez and N. C. Bruce, "Mueller-matrix polarimeter using analysis of the nonlinear voltage–retardance relationship for liquid-crystal variable retarders," *Applied Optics,* 2014.

[32] M. Schulz et al., "Giant intrinsic circular dichroism of prolinol-derived squaraine thin films," *Nat Commun,* 2018.

[33] K. Ramesh and V. Ramakrishnan, "Digital photoelasticity of glass: A comprehensive review," *Optics and Lasers in Engineering,* 2016.

[34] Y. Zhang et al., "Skyrmions based on optical anisotropy for topological encoding," *arXiv preprint arXiv:2508.16483,* 2025.

[35] K. Kikuchi, "Fundamentals of coherent optical fiber communications," *Journal of lightwave technology.*

[36] C. He et al., "A reconfigurable arbitrary retarder array as complex structured matter," *Nature Communications,* 2025.

[37] A. A. Wang et al., "Perturbation-resilient integer arithmetic using optical skyrmions," *Nat. Photon.,* 2025.

[38] A. A. Wang et al., "Generalized skyrmions," *arXiv preprint arXiv:2409.17390,* 2024.

[39] S.-Y. Lu and R. A. Chipman., "Interpretation of Mueller matrices based on polar decomposition.," *Journal of the Optical Society of America,* 1996.

[40] R. Zhang et al., "Elliptical vectorial metrics for physically plausible polarization information analysis," *Advanced Photonics Nexus,* 2025.

[41] L. Deng et al., "A Dual-Modality Imaging Method Based on Polarimetry and Second Harmonic Generation for Characterization and Evaluation of Skin Tissue Structures," *International Journal of Molecular Sciences,* 2023.

[42] T. Novikova and J. C. Ramella-Roman, "Is a complete Mueller matrix necessary in biomedical imaging?," *Optics Letters,* 2022.

[43] J. Chang et al., "Division of focal plane polarimeter-based 3× 4 Mueller matrix microscope: a potential tool for quick diagnosis of human carcinoma tissues.," *Journal of biomedical optics,* 2016.

[44] Z. Karl et al., "Super-resolution dipole orientation mapping via polarization demodulation.," *Light: Science & Applications,* 2016.